\documentstyle[aps,prc,psfig,manuscript]{revtex}

\begin{document}
\bibliographystyle{h-physrev}
%\draft

\title{Hadronic freeze-out following a first order
  hadronization phase transition in ultrarelativistic heavy-ion collisions}

\author{S.A. Bass}
\address{
        Department of Physics\\
        Duke University\\
        Durham, NC, 27708-0305, USA
        }

\author{A. Dumitru}
\address{Department of Physics\\
        Yale University\\
        New Haven, CT, 06520, USA
        }

\author{M. Bleicher, L. Bravina, E. Zabrodin, H. St\"ocker and W. Greiner}
\address{
        Institut f\"ur Theoretische Physik \\
        Johann Wolfgang Goethe Universit\"at\\
        Robert Mayer Str. 8-10\\
        D-60054 Frankfurt am Main, Germany}

\maketitle

\begin{abstract}
We analyze the hadronic freeze-out in ultra-relativistic heavy ion collisions
at RHIC in a transport approach which combines
hydrodynamics for the early, dense, deconfined stage of the reaction 
with a microscopic non-equilibrium model for the later hadronic
stage at which the hydrodynamic equilibrium assumptions are not valid. 
With this ansatz we are able to self-consistently calculate the
freeze-out of the system and determine space-time hypersurfaces for individual
hadron species. 
The space-time domains of the freeze-out for several
hadron species are found to be  actually four-dimensional, 
and differ drastically for the individual hadrons species.
Freeze-out radii distributions are similar
in width for most hadron species, even though the $\Omega^-$ 
is found to be emitted rather close to the phase boundary 
and shows the smallest freeze-out radii and times among all
baryon species.
The total lifetime of the system does not change by more than 10\% when going
from SPS to RHIC energies. 
\end{abstract}

\pagebreak

Ultra-relativistic heavy ion collisions are the only means available
to investigate highly excited dense nuclear matter under controlled laboratory
conditions. In such collisions it is sought to recreate a 
Quark Gluon Plasma (QGP), 
the highly excited state of 
primordial matter which is believed to have existed
shortly after the creation of the universe in 
the Big Bang (for recent reviews on the QGP, 
we refer to \cite{harris96a}). 

Transport theory has been among the most successful approaches applied
to the theoretical investigation of relativistic heavy ion
collisions. 
Microscopic transport models attempt to describe the 
evolution of the heavy ion reaction from some initial state
up to the freeze-out 
of the newly produced particles on the basis of
{\em elementary interactions}.
The basic constituents in such models are either hadrons 
\cite{peilert88a,uqmdref1}
or partons \cite{geiger}. At RHIC
energies, however, both, partonic and hadronic, degrees
of freedom might be equally important and have both to be treated 
explicitly \cite{bass99vni1}.
However, in such microscopic transport models, 
the QG-matter to hadron matter transition, i.e.  
the hadronization stage, has to be modeled in an {\it ad-hoc} 
fashion, whereas  hydrodynamic
approaches~\cite{Bj,clare86a,rischke96a,DumRi,hung98a} 
incorporate this as a phase transition.
This can be done in a consistent way, respecting the laws of thermodynamics
(which is not always the case in microscopic transport models). 
The drawback of hydrodynamics, however, 
is that in the later reaction stages the basic hydrodynamical assumptions
break down. For the freeze-out of the system a decoupling 
(freeze--out) hyper-surface must be specified (or fine-tuned to existing
data).

In this letter, we  use boost-invariant hydrodynamics to model a
first order phase transition from a QGP to a hadronic fluid,
and combine it with a non-equilibrium microscopic
transport calculation for the later, purely hadronic stages of the reaction.
With this ansatz we are able to self-consistently calculate the
freeze-out of the system: no decoupling hypersurface is imposed by hand,
but the space-time points are rather determined by an interplay between
the (local) expansion scalar $\partial u$~\cite{locexp,hung98a}
(where $u$ is the collective flow four-velocity), the relevant elementary
cross sections, and the equation of state (EoS), which actually changes
dynamically as more and more hadron species decouple.

Let us first briefly describe the hydrodynamical model
employed here: For a more detailed discussion we refer to
refs.~\cite{DumRi,dumitru99a,Bassip}.
For simplicity,  boost-invariant longitudinal flow~\cite{Bj} is assumed.
For ultrarelativistic collisions, this should be a reasonable first
approximation in the central rapidity region. Cylindrically
symmetric transverse expansion is superimposed.
For $T>T_C=160$~MeV the well-known 
MIT bag model equation of state \cite{chodos74a} is used,
assuming for simplicity an ideal gas of quarks,
anti-quarks (with masses $m_u=m_d=0$, $m_s=150$~MeV), and gluons.
For $T<T_C$ an ideal hadron gas is employed that includes the complete
hadronic spectrum up to a mass of 2~GeV.
At $T=T_C, (\mu_B=\mu_S=0)$  we require that both pressures are equal, which
fixes the bag constant to $B=380$~MeV/fm$^3$.
By construction the EoS exhibits a first order phase transition (which
is also expected in QCD for the quark-hadron phase transition in the 
case of three quark flavors).

The model reproduces the measured $p_T$- and $m_T$-spectra of hadrons at
the SPS, when assuming that hydrodynamic flow sets in
on the proper time hyperbola $\tau_i=1$~fm/c~\cite{DumRi,dumitru99a}.
This is a value conventionally
assumed in the literature, cf.\ e.g.~\cite{Bj}.
Due to the higher parton density at midrapidity,
thermalization may be reached earlier at RHIC \cite{geiger92a}. 
As in refs.~\cite{DumRi,dumitru99a}, we assume here $\tau_i=R_T/10=0.6$~fm.
The effects of  variations of $\tau_i$ 
and $T_C$
will be discussed
in a future publication~\cite{Bassip}. Moreover, we use the 
initial average energy and baryon densities
$\overline\epsilon(\tau_i)=20$~GeV/fm$^3$ and $\overline\rho(\tau_i)=
2.3\rho_0$, which lead to $dN_B/dy=25$ and $\overline{s}/\overline\rho_B=205$
(a bar symbols an average over the transverse plane).
The initial energy and net baryon densities
are assumed to be distributed in the transverse
plane according to a so-called ``wounded nucleon'' distribution $\propto
\frac{3}{2}\sqrt{1-r_T^2/R_T^2}$, with transverse radius $R_T=6$~fm.
For this set of parameters, the initial transverse energy at midrapidity is
${\rm d}E_T/{\rm d}y=1.3$~TeV. Due to the work performed by the isentropic
expansion, it decreases to 716~GeV on the hadronization hypersurface.
The microscopic treatment of the hadronic dynamics following
hadronization (see below) yields ${\rm d}E_T/{\rm d}y=714$~GeV at
kinetic freeze-out. Thus, the late hadronic evolution at RHIC energy is
not isentropic.

After specifying the initial conditions and the EoS, we determine numerically 
the hydrodynamical solution between the $\tau=\tau_i$ hyperbola and the
hadronization hypersurface, where we apply the Cooper-Frye formula~\cite{CF} to
obtain the hadron spectra. However, in contrast to the usual procedure
we do not integrate over the hypersurface, because we further on also need the
{\em space-time} distribution of hadrons emerging from the mixed phase,
not only their momentum-space distributions~\cite{dumitru99a}.
The ensemble of hadrons thus generated is
then used as 
initial condition for the non-equilibrium microscopic transport model
Ultra-relativistic Quantum Molecular Dynamics (UrQMD) 
\cite{uqmdref1,uqmdref2}. 
The UrQMD model 
contains hadronic (and string) degrees of freedom -- 
all hadronic states can be produced in string decays, s-channel
collisions or resonance decays. Tabulated and parameterized experimental 
cross sections are used when available. Resonance absorption, decays 
and scattering are handled via the principle of detailed balance. 
The UrQMD model has been extensively tested in the SIS, AGS and SPS
energy domain and provides a robust description of hadronic
heavy-ion physics phenomenology. An extensive description of the model, as
well as comparisons with various available data can be found in
\cite{uqmdref1,uqmdref2}.

During the mixed phase the system is either described {\em locally} within
the hydrodynamical framework (as long as a non-zero 
fraction of the fluid in the cell 
consists of quark and gluons) or within the microscopic transport (in the case
of pure hadronic matter). Therefore there exists a time interval 
during the reaction in which both models are applied
in parallel, even though they never refer locally to the same space-time
volume.

\begin{figure}
\centerline{\psfig{figure=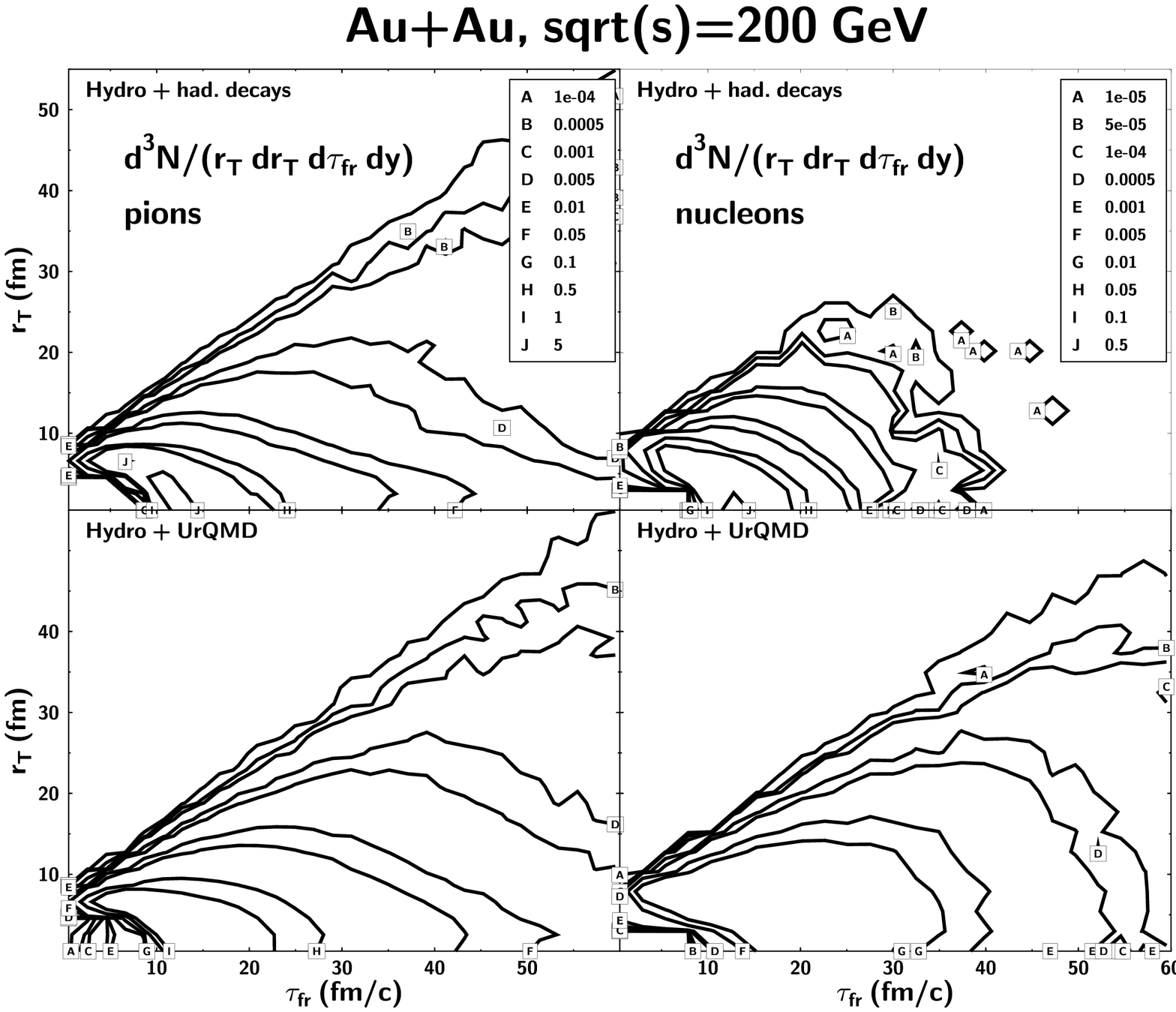,width=5in}}
\caption{\label{focontour} Freeze-out time and transverse radius distribution
d$^3N/(r_T$d$r_T$d$\tau_{fr}$d$y)$ for pions (left column) and protons
(right column).
The top row shows the result for the pure hydro case up to hadronization with
subsequent hadron resonance decays (but without hadronic reinteraction).
The bottom row shows the analogous calculation, but 
with full microscopic hadronic collision
dynamics after the hadronization. The contour lines
have identical binning within each column but differ between the two columns. }
\end{figure}

Let us now turn to the reaction dynamics of central (impact parameter 
$b = 0 $~fm) Au+Au collisions at RHIC energies ($\sqrt{s}=200$~GeV 
per incident colliding nucleon-pair). We start with the 
analysis of the freeze-out hypersurfaces of pions and
nucleons, the most abundant meson and baryon species in the system,
restricting ourselves to the
central rapidity region $y=|y_{CM}|\le 0.5$.
Figure~\ref{focontour} shows the
freeze-out time distributions and the transverse radius distributions
for both, pions and nucleons.
The top row shows the result of a pure hydrodynamical calculation 
up to complete hadronization,
with subsequent hadronic decays,
but without hadronic reinteraction. 
The bottom row
shows the same calculation with full microscopic hadronic dynamics added.
 
The freeze-out characteristics of both, pions and especially 
nucleons, are 
significantly modified due to the hadronic interaction phase. The average
transverse freeze-out radius of the pions changes from 7.8 to 9.5 fm and that
of the protons doubles from 5.4 to 11.3 fm. Their respective
average freeze-out times change from 17.2 to 23.1 fm/c (pions) and from 
11.3 to 25.8 fm/c (protons). 
As the meson multiplicity in the system is fifty times
larger than the baryon multiplicity, baryons propagate through the
relativistic meson gas -- they may act as probes of this highly 
excited meson medium.
Thus, a first estimate of the duration of the hadronic phase is
$\Delta \tau \approx 13$~fm/c. Its transverse spatial 
extent is on the order of $\Delta r_T \approx 6$~fm.

The Hydro+UrQMD model predicts a space-time 
freeze-out picture which is drastically different from that usually 
employed in the hydrodynamical model, e.g.\ in
refs.~\cite{rischke96a,DumRi,hung98a,locexp,AltFO,Cley97}:
Freeze-out here is found to occur
in a {\em four-dimensional} region within the forward
light-cone~\cite{Grassi97}
rather than on a three-dimensional ``hypersurface'' in
space-time~\cite{CF}. Similar results have
also been obtained within other microscopic transport models~\cite{microFO}
when the initial state was not a quark-gluon plasma.
This finding seems to be a generic feature of
such models:  the elementary binary hadron-hadron interactions smear
out the sharp signals to be expected from simple hydro.
This predicted additional fourth dimension
of the freeze-out domain could affect the HBT parameters considerably. 

This does not mean that the {\em momentum-distributions} alone
can not be calculated assuming freeze-out on some effective three-dimensional
hypersurface. (For example, if interactions on the outer side of that
hypersurface are very ``soft'', the single-particle momentum distributions
will not change anymore,
while the two-particle correlator {\em does} change. Thus, the freeze-out
condition, e.g.\ the temperature, as measured by single-particle spectra
and two-particle correlations~\cite{TPFO} needs not be the same.) 

The shapes of the freeze-out hypersurfaces (FOHS)
show broad radial maxima for intermediate 
freeze-out times. Thus, transverse expansion has not developed scaling-flow
(in that case the FOHS would be hyperbolas in the $\tau-r_T$ plane).
Moreover, the hypersurfaces of pions and nucleons, and
their shapes, are distinct from each other
(as also found in \cite{uqmdref1,hung98a,microFO,Pratt98}
at the lower BNL-AGS and CERN-SPS energies). Thus, our
calculation contradicts the ansatz of a unique freeze-out hypersurface
for all hadrons, cf.\ also refs.~\cite{dumitru99a,microFO}.

\begin{figure}
\centerline{\psfig{figure=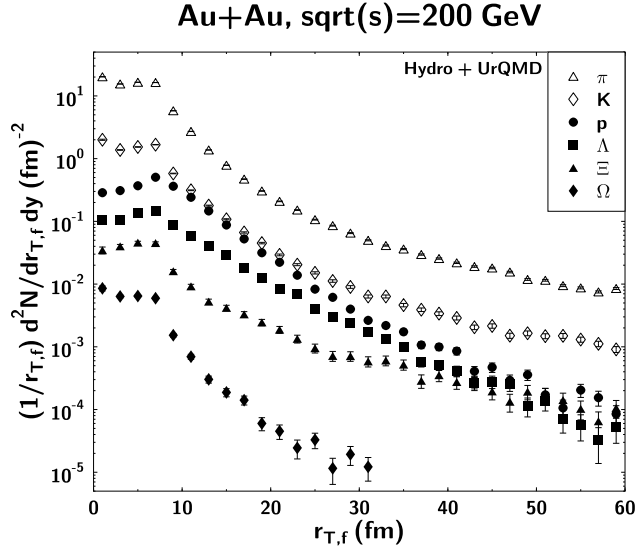,width=3.5in}}
\caption{\label{dndrt} Transverse freeze-out radius distributions
d$^2N/r_{T,f}$d$r_{T,f}$d$y$
for various hadron species. The distributions for
$\pi$, $K$, $p$, $\Lambda$ and $\Xi$ are broad and similar to each other, 
whereas the $\Omega^-$ exhibits a  narrower freeze-out distribution.  }
\end{figure}

Figure~\ref{dndrt} shows the transverse freeze-out radius distributions
for 
$\pi$, $K$, $p$, $\Lambda$, $\Xi$ and $\Omega^-$. 
They are rather broad and similar to each other, 
even though the $\Omega^-$ shows a somewhat narrower freeze-out distribution.
The average transverse freeze-out radii are 9.5~fm for pions, 10.2~fm for
kaons,
11.3~fm for protons, 11.6~fm for Lambda- and Sigma-Hyperons, 14.2~fm for
Cascades, but only 7.3~fm for the $\Omega^-$. The freeze-out of the 
$\Omega^-$ occurs rather close to the phase-boundary~\cite{dumitru99a}, due to
its very small hadronic interaction cross section.
This behavior could be responsible   
for the  experimentally observed
hadron-mass dependence of the inverse slopes 
of the $m_T$-spectra at SPS energies \cite{hecke98a}.  
For the $\Omega^-$, the inverse slope remains
practically unaffected by the purely
hadronic stage of the reaction, due to its small interaction cross section,
while the flow of $p$'s and
$\Lambda$'s increases \cite{dumitru99a}. 
By comparing the  transverse freeze-out radii of the 
hydrodynamical calculation 
(up to hadronization, including subsequent hadronic decays,
but no hadronic reinteractions) with 
the Hydro+UrQMD calculation, which include 
microscopic hadronic dynamics, 
the {\em thickness} $\Delta r_{had}$ 
of the hadronic phase can be estimated by computing the 
difference: $\Delta r_{had} = \langle r_{t,fr}^{\rm Hydro+UrQMD}\rangle -
\langle r_{t,fr}^{\rm Hydro + had. decays} \rangle$. 
These values for $\Delta r_{had}$ are: 1.7~fm for pions, 
3.1~fm for kaons, 5.8~fm for 
protons, lambda- and sigma-hyperons as well as cascades and 2.6~fm for
the $\Omega^-$.

\begin{figure}
\centerline{\psfig{figure=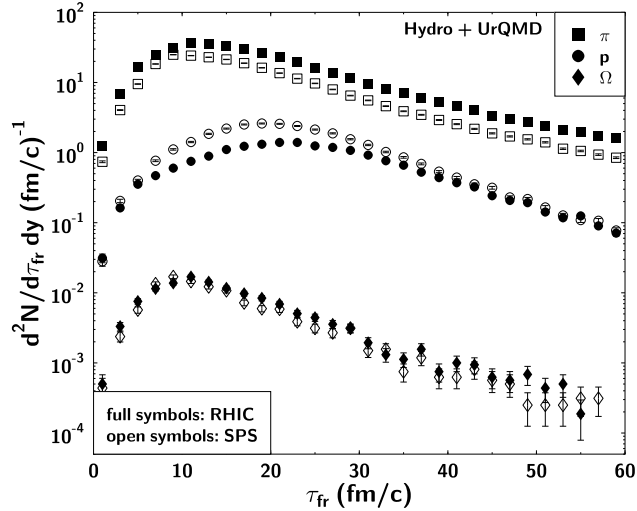,width=3.5in}}
\caption{\label{dndtf} Freeze-out time distributions
d$^2N/$d$\tau_{fr}$d$y$ of
$\pi$, $p$ and $\Omega^-$ for SPS and RHIC.
Apart from the different integral values there is no significant difference
between the RHIC and SPS distributions, i.e.\ the total lifetime of 
the reaction is comparable. }
\end{figure}

Another issue of interest is the predicted significant increase of
the lifetime of the system from SPS to RHIC energies~\cite{rischke96a}.
Figure~\ref{dndtf} shows that in our model, which exhibits a first order
phase transition, there is between SPS and RHIC no difference
in the freeze-out time distributions of
$\pi$, $p$, and $\Omega^-$!
Origin of this prediction is that we
include many more states in the hadronic EoS, which speeds up hadronization
considerably~\cite{DumRi,Cley97}. Furthermore, decays of resonances
(which were not treated in~\cite{rischke96a}) mask the remaining 
small increase of the hadronization time.
Note that the
multistrange $\Omega^-$ baryons freeze out far earlier than all other baryons,
as discussed already previously in the context of figure~\ref{dndrt}.
The {\em duration} of the hadronic reinteraction phase,
$\Delta \tau_{had} = \langle \tau_{fr}^{\rm Hydro+UrQMD}\rangle - 
\langle \tau_{fr}^{\rm Hydro + had. decays} \rangle $
remains nearly unchanged, e.g. at 5.9~fm/c for pions, 8.0~fm/c for kaons,
14.5~fm/c for protons, 15.4~fm/c for hyperons and 8.0~fm/c for the $\Omega^-$.

Note that the lifetime of the prehadronic stage in this approach 
is a factor of  $2-3$ longer than when employing the parton 
cascade model (PCM) \cite{geiger,bass99vni1} for the initial reaction stage. 
It will be interesting to check whether this is related to the 
first-order phase transition built into the EoS which is used here. 
The final transverse freeze-out radii and times (after hadronic rescattering),
however, are very similar in both approaches \cite{bass99vni1}.

\begin{figure}
\centerline{\psfig{figure=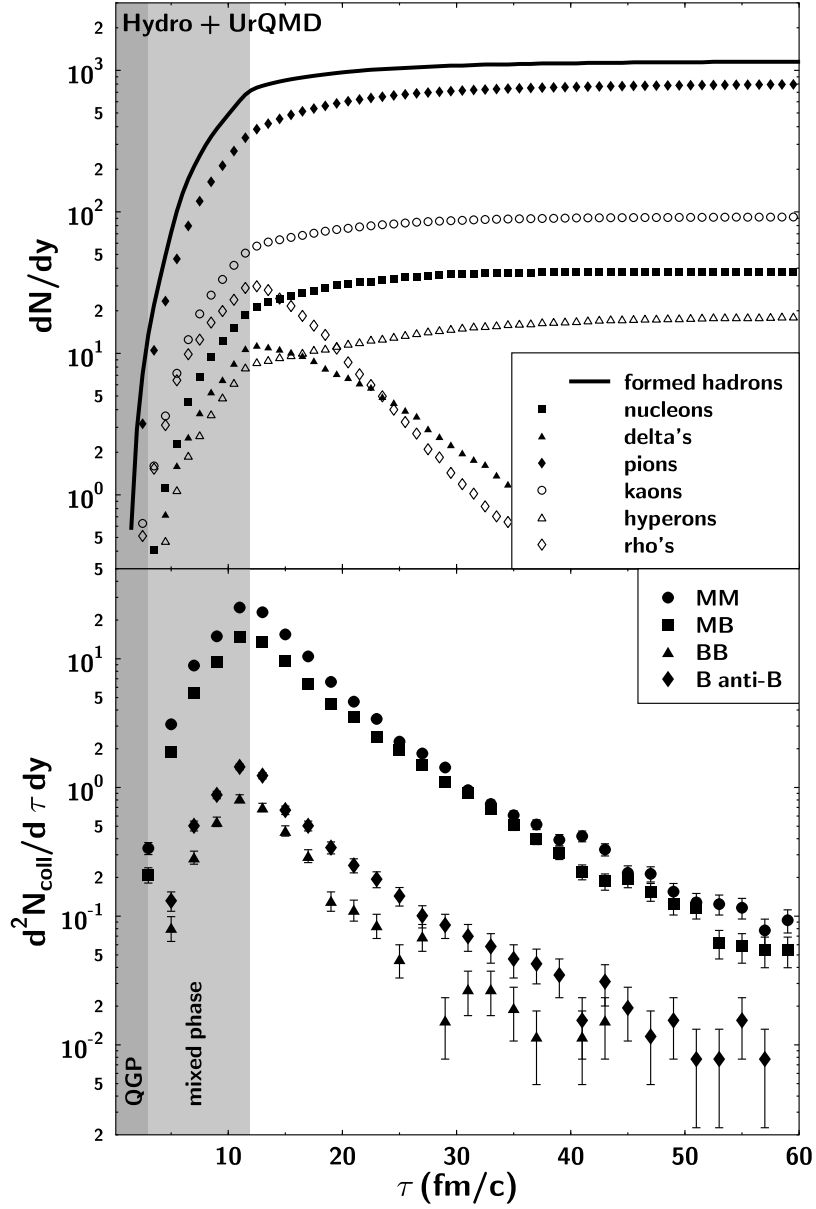,width=4.5in}}
\caption{\label{tevol} Top: time evolution of on-shell hadron
multiplicities (integrated over $r_T$). The dark grey shaded area shows 
the duration of the QGP phase whereas the light grey shaded area depicts the
coexistance phase. 
Bottom: Hadron-hadron collision rates.}
\end{figure}

So far, we have only discussed the 
kinetic freeze-out of individual hadron species,
which is the most precisely determinable freeze-out quantity 
of the system. However, apart from the kinetic freeze-out, 
the chemical freeze-out of the system, which fixes the chemical
composition is of interest. The top frame
of figure~\ref{tevol} shows the time evolution of on-shell hadron
multiplicities. The dark grey shaded area indicates 
the duration of the QGP phase whereas the light grey shaded area depicts the
mixed phase (both averaged over $r_T$; only hadrons that have already
``escaped'' from the mixed phase are shown). Hadronic resonances
are formed and are populated for a long time ($\approx 20$fm/c).
When the mixed phase ceases to exist, the hadron yields have not yet saturated
(even if resonance decays are taken into account). This is due to 
inelastic hadron-hadron collisions.
In particular the yield of anti-protons drops strongly
-- more than 60\% of the baryon-anti-baryon annihilations 
occur after the phase-coexistence 
is over (c.f.\ the lower frame of this figure). 
The yields of all stable hadrons saturate at approximately 25~fm/c.
Only then may the system be viewed as chemically
frozen-out. Since resonance decays have not been included into our estimate
of the saturation time, this number may be viewed as an upper estimate of
the chemical freeze-out time.

By  comparing different final hadron yields resulting from
the hydrodynamical calculation 
(up to hadronization, including subsequent hadronic decays,
but no hadronic reinteractions) to that of 
the Hydro+UrQMD calculation, which includes 
microscopic hadronic dynamics, we can quantify the changes of the 
hadrochemical content due to hadronic rescattering: especially the 
multiplicities of (anti)baryons
vary at least by~10\%, those of protons and antiprotons even up to~30\%
($\pi: +9.3\%,\,\, K: -5\%,\,\, Y: +12\%,\,\, p: -21\%,\,\, \bar{p}: -31\%,
\,\, \bar{Y}: +11\%$).
These changes clearly indicate that in our model 
chemical freeze-out of (anti-)baryons and (anti-)hyperons does not occur
directly at the phase-boundary. Also, note that (unlike in ideal chemical
equilibrium) baryon number is ``shuffled'' from non-strange to strange
baryons.

The bottom frame of figure~\ref{tevol} shows the rates for hadron-hadron 
collisions. Meson-meson (MM) and -- to a lesser extent -- meson-baryon (MB)
interactions dominate the dynamics in the hadronic phase. 
However, the $B\bar B$ collisions outnumber $BB$ reactions, 
in clear contrast to SPS. 
This is a consequence of the fact that the $B\bar B$
annihilation cross sections at small relative momenta increase faster
then the total $BB$ cross section \cite{uqmdref1}. 
In the case of (approximate) baryon-antibaryon symmetry, 
one therefore expects more $B\bar B$
than $BB$ interactions, as seen in figure~\ref{tevol}.

All collision rates
reach their maxima at the end of the mixed phase -- then they decrease
roughly according to a power-law.
After $\approx35$~fm/c, less than one hadron-hadron collision occurs
per unit of time and rapidity -- at this stage the system can be 
considered as kinetically frozen-out. 

In summary, 
we have analyzed the hadronic freeze-out in ultra-relativistic 
heavy ion collisions
at RHIC in a transport approach which combines 
hydrodynamics for the early, dense, deconfined stage of the reaction 
with a microscopic non-equilibrium model for the later hadronic
stage at which the hydrodynamic equilibrium assumptions are not valid anymore. 
Within this approach we have self-consistently calculated the
freeze-out of the hadronic system and
accounted for the collective flow on the
hadronization hypersurface generated by the QGP expansion.

We find that the space-time domains of the freeze-out for the investigated
hadron species are actually four-dimensional, and differ drastically
between the individual hadrons species.

The {\em thickness} of the hadronic phase is found to be between 2~fm and
6~fm, depending on the respective hadron species. Its {\em lifetime} is between
5~fm/c and 13~fm/c, respectively. Freeze-out radii distributions are similar
in width for most hadron species, even though the $\Omega^-$ 
is found to be emitted rather close to the phase boundary 
and shows the smallest freeze-out radii and times among all
baryon species.
The total lifetime of the system does not change by more than 10\% when going
from SPS to RHIC energies. 
Finally, we have found in our model that chemical freeze-out of (anti-)baryons 
does not occur at the phase boundary and
precedes the kinetic freeze-out of the system. 

\acknowledgements
S.A.B.\ is supported in part by the Alexander von Humboldt Foundation
through a Feodor Lynen Fellowship, and in part by DOE grant DE-FG02-96ER40945.
A.D.\ gratefully acknowledges a postdoctoral fellowship by the German
Academic Exchange Service (DAAD) and L.~Bravina is supported
by the Alexander von Humboldt Foundation as a Humboldt Fellow.
S.A.B. acknowledges many helpful and inspiring discussions 
with Berndt M\"uller. A.D. and S.A.B also wish to thank Ulrich Heinz for useful
comments.

%\bibliography{main}

%\newpage

%\newpage

\end{document}